\documentclass[11pt]{article}

\usepackage{amsfonts}
\usepackage{amsmath}
\usepackage{graphicx}
\usepackage{color}
\usepackage{vmargin}

\setpapersize{USletter}
\setmarginsrb{1in}{0.75in}{1in}{0.6in}{0in}{0in}{11pt}{0.4in} 

\begin{document}

\title{
Algorithmic and  Statistical Challenges in 
Modern Large-Scale Data Analysis are the Focus of MMDS 2008
}

\author{
Michael~W.~Mahoney
\and Lek-Heng~Lim
\and Gunnar~E.~Carlsson
}

\date{}
\maketitle

\vspace{0.25in}

The 2008 Workshop on Algorithms for Modern Massive Data Sets (MMDS 2008), 
sponsored by the NSF, DARPA, LinkedIn, and Yahoo!, 
was held at Stanford University, June 25--28. 
The goals of MMDS 2008 were 
(1) to explore novel techniques for modeling and analyzing massive, 
high-dimensional, and nonlinearly-structured scientific and internet data 
sets; 
and (2) to bring together computer scientists, statisticians, 
mathematicians, and data analysis practitioners to promote 
cross-fertilization of ideas.

MMDS 2008 originally grew out of discussions about our vision for 
the next-generation of algorithmic, mathematical, and statistical analysis 
methods for complex large-scale data sets. 
These discussions occurred in the wake of MMDS 2006, which was originally 
motivated by the complementary perspectives brought by the numerical 
linear algebra and theoretical computer science communities to 
matrix algorithms in modern informatics 
applications~\cite{MMDS06summary}.
As with the original 2006 meeting, the MMDS 2008 program generated intense 
interdisciplinary interest: 
with $43$ talks and $18$ poster presentations from a wide spectrum of 
researchers in modern large-scale data analysis, including both senior 
researchers well-established as leaders in their respective fields as well 
as junior researchers promising to become leaders in this new 
interdisciplinary field, the program drew nearly $300$ participants.

\section*{Diverse Approaches to Modern Data Problems}

Graph and matrix problems were common topics for discussion, largely since 
they arise naturally in almost every aspect of data mining, machine learning, 
and pattern recognition.
For example, a common way to model a large social or information network is 
with an \emph{interaction graph model}, $G=(V,E)$, in which nodes in the vertex 
set $V$ represent ``entities'' and the edges (whether directed, undirected, 
weighted or unweighted) in the edge set $E$  represent ``interactions'' 
between pairs of entities. 
Alternatively, these and other data sets can be modeled as 
matrices, since an $m \times n$ real-valued matrix $A$ provides a natural 
structure for encoding information about $m$ objects, each of which is 
described by $n$ features.  
Due to their large size, their extreme sparsity, and their complex and often 
adversarial noise properties, data graphs and data matrices arising in
modern informatics applications present considerable challenges and 
opportunities for interdisciplinary research. 
These algorithmic, statistical, and mathematical challenges 
were the focus of MMDS~2008.

It is worth emphasizing the very different perspectives that have 
historically been brought to such problems.
For example, a common view of the data in a database, in particular 
historically  among computer scientists interested in data mining and 
knowledge discovery, has been that the data are an accounting or a 
record of everything that happened in a particular setting.
For example, the database might consist of all the customer transactions 
over the course of a month, or it might consist of all the friendship 
links among members of a social networking site.
From this perspective, the goal is to tabulate and process the data at hand 
to find interesting patterns, rules, and associations.
An example of an association rule is the proverbial ``People who buy beer
between $5$ p.m.\ and $7$ p.m.\ also buy diapers at the same time.''
The performance or quality of such a rule is judged by the fraction of the 
database that satisfies the rule exactly, which then boils down to the 
problem of finding frequent itemsets.
This is a computationally hard problem, and much algorithmic work has been 
devoted to its exact or approximate solution under different models of data 
access.

A very different view of the data, more common among statisticians, is one of 
a particular random instantiation of an underlying process describing 
unobserved patterns in the world.
In this case, the goal is to extract information about the world from the 
noisy or uncertain data that is observed.
To achieve this, one might posit a model: 
$ \mathit{data} \sim F_{\theta} $ and 
$ \operatorname*{mean}(\mathit{data}) = g(\theta) $,
where $F_{\theta}$ is a distribution that describes the random variability
of the data around the deterministic model $g(\theta)$ of the data.
Then, using this model, one would proceed to analyze the data to make 
inferences about the underlying processes and predictions about future 
observations. 
From this perspective, modeling the noise component or variability well is as 
important as modeling the mean structure well, in large part since 
understanding the former is necessary for understanding the quality of 
predictions made.
With this approach, one can even make predictions about events that have 
yet to be observed.
For example, one can assign a probability to the event that a given user 
at a given web site will click on a given advertisement presented at a 
given time of the day, even if this particular event does not exist in the 
database.

The two perspectives need not be incompatible.
For example, statistical and probabilistic ideas are central to much of the 
recent work on developing improved approximation algorithms for matrix 
problems;
otherwise intractable optimization problems on graphs and networks yield to 
approximation algorithms when assumptions are made about the network 
participants; 
much recent work in machine learning draws on ideas from both areas; and
in boosting, a statistical technique that fits an additive model by 
minimizing an objective function with a method such as gradient descent, the 
computation parameter, i.e., the number of iterations, also serves as 
a regularization parameter.

Given the diversity of possible perspectives, MMDS 2008 was 
organized loosely 
around six hour-long tutorials that introduced participants to the 
major themes of the workshop.

\section*{Large-Scale Informatics: Problems, Methods, and Models}

On the first day of the workshop, participants heard tutorials by Christos 
Faloutsos of Carnegie Mellon University and Edward Chang of Google Research, 
in which they presented an overview of tools and applications in modern 
large-scale data analysis.

Faloutsos began his tutorial on ``Graph mining: laws, generators and tools''
by motivating the problem of data analysis on graphs.
He described a wide range of applications in which graphs arise naturally, 
and he reminded the audience that large graphs that arise in modern 
informatics applications have structural properties that are very different 
from traditional Erd\H{o}s-R\'{e}nyi random graphs.
For example, due to subtle correlations, statistics such as degree 
distributions and eigenvalue distributions exhibit heavy-tailed behavior.

Although these structural properties have been studied extensively in recent
years and have been used to develop numerous well-publicized models, 
Faloutsos also described empirically-observed properties that are not 
well-reproduced by existing models.
As an example, most models predict that over time the graph should become 
sparser and the diameter should grow as $O(\log N)$ or perhaps 
$O(\log \log N)$, where $N$ is the number of nodes at the current time step, 
but empirically it is often observed that the networks densify over time and 
that their diameter shrinks.
To explain these phenomena, Faloutsos described a model based on Kronecker 
products and also a model in which edges are added via an iterative ``forest 
fire'' burning mechanism.
With appropriate choice of parameters, both models can be made to reproduce 
a much wider range of static and dynamic properties than can previous 
generative models.

Building on this modeling foundation, Faloutsos spent much of his talk 
describing several graph mining applications of recent and ongoing interest:
methods to find nodes that are central to a group of individuals; 
applications of the Singular Value Decomposition and recently-developed 
tensor methods to identifying anomalous patterns in time-evolving graphs; 
modeling information cascades in the blogosphere as virus propagation; and 
novel methods for fraud detection.

Edward Chang described other developments in web-scale data analysis in his 
tutorial on ``Mining large-scale social networks: challenges and scalable 
solutions.''
After reviewing emerging applications---such as social network analysis and 
personalized information retrieval---that have arisen as we make the transition from 
Web 1.0 (links between pages and documents) to Web 2.0 (links between 
documents, people, and social platforms), Chang covered 
in detail
four applications 
his team recently parallelized:
spectral clustering for network analysis, 
frequent itemset mining, 
combinatorial collaborative filtering, 
and parallel Support Vector Machines (SVMs) for personalized search.
In all these cases, he emphasized that the main performance requirements 
were ``scalability, scalability, scalability.'' 

Modern informatics applications like web search afford easy 
parallelization---e.g., the overall index can be partitioned such 
that even a single query can use multiple processors.
Moreover, the peak performance of a machine is less important than the 
price-performance ratio.
In this environment, scalability up to petabyte-sized data often means 
working in a software framework like MapReduce or Hadoop that supports 
data-intensive distributed computations running on large clusters of 
hundreds, thousands, or even hundreds of thousands of commodity computers. 
This differs substantially from the scalability issues that arise in 
traditional applications of interest in scientific computing. 
A recurrent theme of Chang was that an algorithm that is expensive in 
floating point cost but readily parallelizable is often a better choice than 
one that is less expensive but non-parallelizable.

As an example, although SVMs are widely-used, largely due to their empirical 
success and attractive theoretical foundations, they suffer from well-known 
scalability problems in both memory use and computational time.
To address these problems, Chang described a Parallel SVM algorithm.
This algorithm reduces memory requirements by performing a 
\textit{row-based} 
Incomplete Cholesky Factorization (ICF) and by loading only essential data 
to each of the parallel machines; and
it reduces computation time by intelligently reordering computational steps 
and by performing them on parallel machines. 
Chang noted that the traditional \textit{column-based} ICF is better for 
the single machine setting, but it cannot be parallelized as well across 
many machines.

\section*{Algorithmic Approaches to Networked Data}

Milena Mihail of the Georgia Institute of Technology described algorithmic 
perspectives on developing better models for data in her tutorial 
``Models and algorithms for complex networks.'' 
She noted that in recent years a rich theory of power law random graphs, 
i.e., graphs that are random conditioned on a specified input power law 
degree distribution, has been developed.
With the increasingly wide range of large-scale social and information 
networks that are available, however, generative models that are 
structurally or syntactically more flexible are increasingly necessary.
Mihail described two such extensions: 
one in which semantics on nodes is modeled by a feature vector and edges are 
added between nodes based on their semantic proximity; and 
one in which the phenomenon of associativity/disassociativity is modeled by 
fixing the probability that nodes of a given degree $d_i$ tend to link to 
nodes of degree $d_j$.

By introducing a small extension in the parameters of a generative model, of 
course, one can observe a large increase in the observed properties of 
generated graphs.
This observation raises interesting statistical questions about model 
overfitting, and it argues for more refined and systematic methods of 
model parameterization.
This observation also leads to new algorithmic questions that were the 
topic of Mihail's talk.

An algorithmic question of interest in the basic power law random graph 
model is the following: given as input an $N$-vector specifying a degree 
sequence, determine whether there exists a graph with that degree sequence, 
and, if so, efficiently generate one (perhaps approximately uniformly randomly 
from the ensemble of such graphs).
Such realizability problems have a long history in graph theory and 
theoretical computer science.
Since their solutions are intimately related to the theory of graph 
matchings, many generalizations of the basic problem can be addressed in a 
strict theoretical framework.
For example, motivated by associative/disassociative networks, Mihail 
described recent progress on the Joint-Degree Matrix Realization Problem:
given a partition of the node set into classes of vertices of the same 
degree, a vector specifying the degree of each class, and a matrix 
specifying the number of edges between any two classes, determine whether
there exists such a graph, and if so construct one. 
She also described extensions of this basic problem to connected graphs, to 
finding minimum cost realizations, and and to finding a random graph 
satisfying those basic constraints.

\section*{The Geometric Perspective: Qualitative Analysis of Data}

A very different perspective was provided by Gunnar Carlsson of Stanford 
University, who gave an overview of geometric and topological approaches 
to data analysis in his tutorial ``Topology and data.''
The motivation underlying these approaches is to provide insight into the
data by imposing a geometry on it.
Whereas in certain applications, such as in physics, the studied phenomena 
support clean explanatory theories which define exactly the metric to 
use to measure the distance between pairs of data points, in most MMDS 
applications this is not the case.
For instance, the Euclidean distance between DNA 
expression profiles in high-throughput microarray experiments may or may not
capture a 
meaningful notion of distance between genes.
Similarly, although a natural geodesic distance is associated with 
any graph, the sparsity and noise properties of social and information 
networks means that this is not a particularly robust notion of 
distance in practice.

Part of the problem is thus to define useful metrics---in particular since 
applications such as clustering, classification, and regression often depend 
sensitively on the choice of metric---and two design goals have recently 
emerged.
First, don't trust large distances---since distances are often constructed 
from a similarity measure, small distances reliably represent similarity 
but large distances make little sense.
Second, trust small distances only a bit---after all, similarity 
measurements are still very noisy.
These ideas have formed the basis for much of the work on Laplacian-based
non-linear dimensionality reduction, i.e., manifold-based, methods 
that are currently popular in harmonic analysis and machine learning. 
More generally, they suggest the design of analysis tools that are robust
to stretching and shrinking of the underlying metric, particularly
in applications such as visualization in which qualitative properties, such 
as how the data are organized on a large scale, are of interest.

Much of Carlsson's tutorial was occupied by describing these analysis tools 
and their application to natural image statistics and data visualization.
Homology is the crudest measure of topological properties, capturing 
information such as the number of connected components, whether the data 
contain holes of various dimensions, etc.
Importantly, although the computation of homology is not feasible for 
general topological spaces, in many cases the space can be modeled in terms 
of simplicial complexes, in which case the computation of homology boils 
down to the linear algebraic computation of the Smith normal form of 
certain data-dependent matrices.
Carlsson also described \emph{persistent homology}, an extension of the basic 
idea in which parameters such as the number of nearest neighbors, error 
parameters, etc., can be varied.
A ``bar code signature'' can then be associated with the data set.
Long segments in the bar code indicate the presence of a homology class 
which persists over a long range of parameters values.
This can often be interpreted as corresponding to large-scale geometric
 features in the data, while shorter segments can be interpreted as noise.

\section*{Statistical and Machine Learning Perspectives}

Statistical and machine learning perspectives on MMDS were the subject of a 
pair of tutorials by Jerome Friedman of Stanford University and Michael 
Jordan of the University of California at Berkeley.
Given a set of measured values of attributes of an object, 
$\mathbf{x}=(x_1,x_2,\dots,x_n)$, the basic predictive or machine learning 
problem is to predict or estimate the unknown value of another attribute $y$.
The quantity $y$ is the ``output'' or ``response'' variable, and 
$\{x_1,x_2,\dots,x_n\}$ are the ``input'' or ``predictor'' variables.
In regression problems, $y$ is a real number, while in 
classification problems, $y$ is a member of a discrete set of 
unorderable categorical values (such as class labels).
In either case, this can be viewed as a function estimation problem---the 
prediction takes the form of a function $\hat{y}=F(\mathbf{x})$ that maps a 
point $\mathbf{x}$ in the space of all joint values of the predictor variables 
to a point $\hat{y}$ in the space of response variables, and the goal is to 
produce an $F(\cdot)$ that minimizes a loss criterion.

In his tutorial, ``Fast sparse regression and classification,'' Friedman 
began with the common assumption of
a linear model, in which 
$F(\mathbf{x})=\sum_{j=1}^{n}a_jx_j$ is modeled as a linear combination of the 
$n$ basis functions.
Unless the number of observations is much much larger than $n$, however, 
empirical estimates of the loss function exhibit high variance.
To make the estimates more regular, one typically considers a 
constrained or penalized optimization problem
$$
\hat{\mathbf{a}}(\lambda) 
   = \mbox{argmin}_\mathbf{ a} \hat{L}(\mathbf{ a}) + \lambda P_{\gamma}(\mathbf{ a}) ,
$$
where $\hat{L}(\cdot)$ is the empirical loss and $P_{\gamma}(\cdot)$ is a
penalty term.
The choice of an appropriate value for the regularization parameter 
$\lambda$ is a classic model selection problem, for which cross validation 
can be used.
The choice for the penalty depends on what is known or assumed about the 
problem at hand.
A common choice is 
$P_{\gamma}(\mathbf{ a}) 
   = \lVert\mathbf{ a}\rVert_{\gamma}^{\gamma}=\sum_{j=1}^{n}|a_j|^{\gamma}$.
This interpolates between the subset selection problem ($\gamma=0$) and
ridge regression ($\gamma=2$) and includes the well-studied lasso 
($\gamma=1$).
For $\gamma \le 1$, sparse solutions (which are of interest due to parsimony 
and interpretability) are obtained, and for $\gamma \ge 1$, the penalty is
convex.

Although one could choose an optimal $(\lambda,\gamma)$ by cross validation, 
this can be prohibitively expensive, even when the loss and penalty are 
convex, due to the need to perform computations at a large number of 
discretized pairs.
In this case, path seeking methods have been studied.
Consider the path of optimal solutions 
$\{\hat{\mathbf{a}}(\lambda):0\le\lambda\le\infty \}$, which is a one-dimensional
curve in the parameter space $\mathbb{R}^{n}$.
If the loss function is quadratic and the penalty function is piecewise 
linear, e.g., with the lasso, then the path of optimal solutions is 
piecewise linear, and homotopy methods can be used to generate the full 
path in time that is not much more than that needed to fit a single model at
a single parameter value.
Friedman described a generalized path seeking algorithm, which solves 
this problem for a much wider range of loss and penalty functions (including 
some non-convex functions) very efficiently. 

Jordan, in his tutorial ``Kernel-based contrast functions for sufficient 
dimension reduction,'' considered the dimensionality reduction problem in a 
supervised learning setting.
Methods such as Principal Components Analysis, Johnson-Lindenstrauss  
techniques, and recently-developed Laplacian-based non-linear methods are
often used, but their applicability is limited since, e.g., the axes 
of maximal discrimination between two the classes may not align well with 
the axes of maximum variance.
Instead, one might hope that there exists a low-dimensional subspace $S$ of 
the input space $X$ which can be found efficiently and which retains the 
statistical relationship between $X$ and the response space $Y$.
Conventional approaches to this problem of Sufficient Dimensionality 
Reduction (SDR) make strong modeling assumptions about the distribution of 
the covariate $X$ and/or the response $Y$.
Jordan considered a semiparametric formulation, where the conditional 
distribution $p(Y \mid X)$ is treated nonparametrically and the goal is 
estimate 
the parameter $S$.
He showed that this problem could be formulated in terms of conditional 
independence and that it could be evaluated in terms of operators on 
Reproducing Kernel Hilbert Spaces (RKHSs).

Recall that claims about the independence between two random variables can 
be reduced to claims about correlations between them by considering 
transformations of the random variables: 
$X_1$ and $X_2$ are independent if and only if 
$$
\max_{h_1,h_2\in\mathcal{H}} \mbox{Corr}(h_1(X_1),h_2(X_2))=0
$$
for a suitably rich function space $\mathcal{H}$.
If $\mathcal{H}$ is $L_2$ and thus contains the Fourier basis, this reduces 
to a well-known fact about characteristic functions.
More interesting from a computational perspective---recall that by the 
``reproducing'' property, function evaluation in a RKHS reduces to an inner 
product---this also holds for suitably rich RKHSs.
This use of RKHS ideas to solve this SDR problem cannot be viewed as a 
kernelization of an underlying linear algorithm, as is typically the 
case when such ideas are used (e.g., with SVMs) to provide basis 
expansions for regression and classification. 
Instead, this is an example of how RKHS ideas provide algorithmically 
efficient machinery to optimize a much wider range of statistical 
functionals of interest.

\section*{Conclusions and Future Directions}

In addition to other talks on 
the theory of data algorithms, 
machine learning and kernel methods, 
dimensionality reduction and graph partitioning methods, 
and co-clustering and other matrix factorization methods, 
participants heard about a wide variety of data applications, including
movie and product recommendations;
predictive indexing for fast web search; 
pathway analysis in biomolecular folding;
functional MRI, high-resolution terrain analysis, and 
galaxy classification;
and other applications in computational geometry, computer graphics, 
computer vision, and manifold learning.
(We even heard about using approximation algorithms in a novel manner to 
probe the community structure of large social and information networks to 
test the claim that such data are even consistent with the manifold 
hypothesis---they clearly are not.)
In all these cases, scalability was a central issue---motivating discussion 
of external memory algorithms, novel computational paradigms like MapReduce, 
and communication-efficient linear algebra algorithms.
Interested readers are invited to visit the conference website, 
\texttt{http://mmds.stanford.edu}, where the presentations from all 
speakers can be found.

The feedback we received made it clear that MMDS has struck a strong 
interdisciplinary chord.
For example, nearly every statistician commented on the desire for
more statisticians at the next MMDS; 
nearly every scientific computing researcher told us they 
wanted more data-intensive scientific computation at the next MMDS; 
nearly every practitioner from an application domain 
wanted more applications at the next MMDS; 
and nearly every theoretical computer scientist said they wanted more of 
the same.
There is a lot of interest in MMDS as a developing interdisciplinary 
research area at the interface between computer science, statistics, 
applied mathematics, and scientific and internet data applications. 
Keep an eye out for future MMDSs!

\vspace{0.25in}
\noindent
\textbf{Acknowledgments}

\noindent
The authors are grateful to the numerous individuals (in particular, Mayita 
Romero, Victor Olmo, and David Gleich) who provided assistance prior to and 
during MMDS 2008;
to Diane Lambert for providing an interesting perspective on these problems 
that we incorporated here;
and to each of the speakers, poster presenters, and other participants, 
without whom MMDS 2008 would not have been such a success.

\vspace{0.25in}
\noindent
\emph{Michael Mahoney (\texttt{mmahoney@cs.stanford.edu}) is a research 
scientist in the Department of Mathematics at Stanford University.
Lek-Heng Lim (\texttt{lekheng@math.berkeley.edu}) is a Charles Morrey 
assistant professor in the Department of Mathematics at University of 
California, Berkeley.
Gunnar Carlsson (\texttt{gunnar@math.stanford.edu}) is a professor in the 
Department of Mathematics at Stanford University.  }

\end{document}